\documentclass[%
 reprint,
 amsmath,amssymb,
 aps,
]{revtex4-2}

\usepackage{graphicx}
\usepackage{dcolumn}
\usepackage{bm}

\usepackage{slashed}
\usepackage{float}
\usepackage{tikz}
\usepackage{tikz-feynman}

\begin{document}


\title{Magnetic Catalysis of charmonium in the vector channel}

\author{C. A. Dominguez}
 \affiliation{Centre for Theoretical and Mathematical Physics, and Department of Physics, University of Cape Town, Rondebosch 7700, South Africa.}
\author{L. A. Hern\'andez}%
\affiliation{Departamento de F\'isica, Universidad Aut\'onoma Metropolitana-Iztapalapa, Avenida San Rafael Atlixco 186, Ciudad de México 09340, Mexico.}%
\author{Michael Koning}
\affiliation{Centre for Theoretical and Mathematical Physics, and Department of Physics, University of Cape Town, Rondebosch 7700, South Africa.}%


\begin{abstract}
We investigate the impact of an external magnetic field on the vector charmonium system within the framework of Hilbert moment QCD sum rules. By incorporating magnetic corrections to the perturbative contributions of the QCD sector, we analyze the behavior of the hadronic parameters of the $J/\psi$ resonance---namely, its continuum threshold $s_0$, decay constant $f_V$, width $\Gamma_V$, and its mass $M_V$, as functions of the magnetic field strength. Our results show that $s_0$ and $f_V$ increase monotonically, while $\Gamma_V$ decreases significantly and $M_V$ remains essentially constant. These behaviors indicate a strengthening of the hadronic state in the presence of a magnetic field, consistent with the phenomenon of magnetic catalysis. Although magnetic catalysis has traditionally been associated with light-quark systems via chiral symmetry breaking, our results demonstrate that similar effects persist in the heavy-quark sector, despite the absence of chiral dynamics.
\end{abstract}

\maketitle


\section{\label{sec1} Introduction}

The study of the behavior of strongly interacting matter under external conditions has become a cornerstone in understanding fundamental aspects of QCD, such as confinement and chiral symmetry breaking. In particular, the influence of strong magnetic fields on hadronic systems has attracted considerable attention in recent years, both from the lattice QCD community and from effective field-theory approaches~\cite{Miransky:2015ava,Adhikari:2024bfa,Hattori:2023egw}. One of the most remarkable phenomena arising in this context is magnetic catalysis~\cite{Gusynin:1994re,Gusynin:1995nb,DElia:2011koc,Bali:2012zg}, which refers to the enhancement of chiral symmetry breaking induced by an external magnetic field. This effect has been widely studied for systems composed of light quarks, where the chiral condensate plays a central role as an order parameter~\cite{Miransky:2002rp,Fukushima:2012kc,Kamikado:2013pya,Mueller:2014tea,Mueller:2015fka,Hattori:2015aki,Ballon-Bayona:2020xtf}.

In contrast, for systems made of heavy quarks, such as charmonium and bottomonium and other types of resonances, chiral symmetry is explicitly broken due to the large current quark masses. Nevertheless, these systems remain highly sensitive to the surrounding medium and external probes, offering an excellent opportunity to explore non-trivial QCD dynamics beyond the light-quark sector. In particular, the interaction of heavy quarkonia with magnetic fields is of growing interest, not only in connection with heavy-ion collisions, where transient magnetic fields of enormous strength are produced~\cite{Skokov:2009qp,Brandenburg:2021lnj,STAR:2023jdd}, but also due to their potential as clean probes of medium modifications~\cite{Adhikari:2024bfa}.

In this work, we investigate the impact of a constant and uniform magnetic field on the vector charmonium system, focusing on the possible realization of magnetic catalysis in the heavy-quark sector. Our analysis is performed within the framework of Hilbert moment QCD sum rules~\cite{Colangelo:2000dp,Dominguez:2018zzi}, a robust non-perturbative approach that relates QCD dynamics at short distances to hadronic parameters via analyticity and dispersion relations, which has been successful in studying the properties of charmonium and bottomonium systems at finite temperature~\cite{Dominguez:2009mk,Dominguez:2010mx,Dominguez:2013fca}. We compute the changes induced by the magnetic field on hadronic parameters such as the width and decay constant of the $J/\psi$ resonance, and the phenomenological continuum threshold $s_0$, serving as a phenomenological order parameter for deconfinement~\cite{Carlomagno:2019wlh}.

The main novelty of this study lies in the extension of the concept of magnetic catalysis to heavy-quark systems, where no chiral condensate is available to drive the phenomenon. Instead, we examine whether the hadronic parameters extracted from QCD sum rules display a modification, indicative of magnetic catalysis, as the magnetic field increases. Our results show a consistent and non-trivial behavior compatible with magnetic catalysis in this regime, suggesting that the phenomenon may be more general than previously thought and not restricted to light-quark dynamics.

This work is organized as follows. In Section~\ref{sec2}, we present in general the formalism of the Hilbert moment QCD sum rules, adapted to the vector charmonium channel. In Section~\ref{sec3}, we include the magnetic effects in the Hilbert moment QCD sum rules. We detail the input parameters, the construction of the Hilbert moments, and the magnetic corrections to the two-point QCD correlation function. Finally, Section~\ref{sec4} is devoted to the analysis and discussion of the results.

\section{Hilbert Moment QCD Sum Rules for the Vector Charmonium Channel \label{sec2}}

The Hilbert moment QCD sum rule method provides a powerful framework for investigating hadronic properties from the fundamental principles of quantum field theory. This approach begins with the two-point correlation function in the charmonium vector channel, defined as
\begin{equation}
\Pi_{\mu\nu}(q) = i \int d^4x \, e^{i q \cdot x} \langle 0 | T \{ j_\mu(x) j_\nu^\dagger(0) \} | 0 \rangle,
\label{currentcorrelator}
\end{equation}
where the current \( j_\mu(x) = \bar{c}(x) \gamma_\mu c(x) \) describes charm quarks in the vector channel. Owing to Lorentz symmetry and gauge invariance, this tensor correlator in vacuum can be decomposed as
\begin{equation}
\Pi_{\mu\nu}(q) = \left( q_\mu q_\nu - q^2 g_{\mu\nu} \right) \Pi(q^2),
\label{correntcorrelatortransverse}
\end{equation}
so that all the physical information is encoded in the scalar function $\Pi(q^2)$. This function can be analytically continued to the complex $q^2$ plane, where it develops a branch cut along the positive real axis. On this axis, the correlator is determined by hadronic degrees of freedom, while in the rest of the complex plane it is governed by QCD degrees of freedom. For sufficiently large $|q^2|$, perturbative QCD becomes applicable. The function satisfies the dispersion relation
\begin{equation}
\Pi(q^2)|_{\mathrm{QCD}} = \frac{1}{\pi} \int_0^\infty \frac{\text{Im}\,\Pi(s)|_{\mathrm{HAD}}}{s - q^2 - i\epsilon} ds + \text{subtractions},
\label{relationPiandspectralfunction}
\end{equation}
where the subtraction constants are eliminated by taking successive derivatives with respect to $Q^2 = -q^2$, which leads to the definition of the QCD Hilbert moment
\begin{align}
\varphi_N(Q^2)|_{\mathrm{QCD}} & \equiv \frac{(-1)^N}{N!} \left( \frac{d}{dQ^2} \right)^N \Pi(-Q^2)|_{\mathrm{QCD}} \label{QCDhilbertmoment} \\
 & = \varphi_N(Q^2)|_{\mathrm{pQCD}}+\varphi_N(Q^2)|_{\mathrm{npQCD}}, \nonumber
\end{align}
where $N$ is a positive integer, and $\varphi_N(Q^2)|_{\mathrm{pQCD}}$ and $\varphi_N(Q^2)|_{\mathrm{npQCD}}$ denote the perturbative and nonperturbative QCD contributions, respectively. On the hadronic side, the corresponding Hilbert moment is defined as
\begin{equation}
\varphi_N(Q^2)\big|_{\mathrm{HAD}} \equiv \frac{1}{\pi} \int_0^{\infty} \frac{\mathrm{Im} \, \Pi(s)\big|_{\mathrm{HAD}}}{(s + Q^2)^{N+1}} ds.
\label{HilbertmomentsHAD}
\end{equation}
According to the quark-hadron duality ansatz, the same moments can be evaluated from two complementary perspectives: on the QCD side, through perturbative and nonperturbative contributions, and on the hadronic side, by modeling the spectral function in terms of resonances and the continuum. Therefore, one can write 
\begin{equation}
   \varphi_N (Q^2)\big|_{\mathrm{QCD}}=\varphi_N(Q^2)\big|_{\mathrm{HAD}}.
   \label{quarkhadronduality}
\end{equation}
In their current form, these Hilbert moments do not explicitly involve the continuum threshold parameter $s_0$. Since this parameter plays a crucial role in characterizing the confinement properties of the hadronic state, it must be incorporated explicitly. For this reason, we define
\begin{equation}
    I_N(Q^2)\equiv \frac{1}{\pi} \int_{s_0}^{\infty} \frac{\mathrm{Im} \, \Pi(s)\big|_{\mathrm{HAD}}}{(s + Q^2)^{N+1}} ds,
    \label{subtractionTerm}
\end{equation}
which allows us to introduce the ``subtracted'' Hilbert moments. On the QCD side, the subtracted Hilbert moment is defined as
\begin{equation}
\Phi_N(Q^2)\big|_{\mathrm{QCD}} \equiv \varphi_N (Q^2)|_{\mathrm{QCD}} - I_N(Q^2), 
\label{substractedHilbertmomentQCD}
\end{equation}
while on the hadronic side, it is given by
\begin{align}
 \Phi_N(Q^2)|_{\mathrm{HAD}}&\equiv\varphi_N (Q^2)|_{\mathrm{HAD}} - I_N(Q^2) \nonumber \\
 &= \frac{1}{\pi} \int_{0}^{s_0} \frac{\mathrm{Im} \, \Pi(s)\big|_{\mathrm{HAD}}}{(s + Q^2)^{N+1}} ds.
 \label{substractedHilbertmomentHAD}
\end{align}
The statement of quark-hadron duality in Eq.~(\ref{quarkhadronduality}) can be extended to the subtracted Hilbert moments, leading to
\begin{equation}
   \Phi_N (Q^2)|_{\mathrm{QCD}}=\Phi_N(Q^2)\big|_{\mathrm{HAD}}.
   \label{quarkhadrondualitySubtracted}
\end{equation}

The hadronic spectral function is a key ingredient in Eq.~(\ref{quarkhadrondualitySubtracted}), as it enters both the subtracted hadronic Hilbert moment and the subtraction term required to compute the subtracted QCD Hilbert moment.  In the charmonium vector channel, the dominant low-energy contribution to the hadronic spectral function arises from the $J/\psi$ meson, while the continuum is modeled using perturbative QCD above the continuum threshold $s_0$. In the zero-width approximation, the hadronic spectral function is written as
\begin{align}
\mathrm{Im} \, \Pi(s)\big|_{\mathrm{HAD}} &= 2\pi f_V^2 \delta(s - M_V^2) \Theta(s_0 - s) \nonumber \\
&+ \mathrm{Im} \, \Pi(s)\big|_{\mathrm{PQCD}} \Theta(s - s_0).
\label{ImPiHAD}
\end{align}
When a finite width is taken into account, the contribution to the low-energy region of the hadronic spectral function is described by the Breit-Wigner parameterization, such that
\begin{equation}
2\pi f_V^2 \delta(s - M_V^2) \to  \frac{2f_V^2 M_V \Gamma_V}{(s-M_V^2)^2+M_V^2\Gamma_V^2},    
\end{equation}
where $\Gamma_V$ denotes the total width of the $J/\psi$.

In practice, the QCD side consists of the perturbative contribution, the gluon condensate (from the nonperturbative contribution), and corrections induced by external effects such as a magnetic field. The parameter $Q^2$ is chosen as a fixed positive quantity to ensure both the convergence of the operator product expansion and the suppression of higher-mass states. The Hilbert moments can then be combined to construct a system of equations that allows the extraction of the hadronic parameters $s_0$, $M_V$, $\Gamma_V$, and $f_V$, which correspond to the continuum threshold, the mass, the total width, and the coupling (leptonic decay constant), respectively. The approach we follow employs the same set of equations used in the study of charmonium at finite temperature~\cite{Dominguez:2009mk}. We divide these equations into two sets:  the first consists of three equations consistent with the finite-width treatment, while the second is a single equation derived in the zero-width approximation. As a result, these two sets of equations use a different momentum parameter, so that the same value for any given hadronic parameter can be used in all of the equations. These sum rule equations, which we refer to as the Hilbert moment QCD sum rules, are given by
\begin{subequations}\label{QCDsumrules}
\begin{align}
\frac{\Phi_1(Q^2)\big|_{\mathrm{QCD}}}{\Phi_2(Q^2)\big|_{\mathrm{QCD}}} &= \frac{\Phi_2(Q^2)\big|_{\mathrm{QCD}}}{\Phi_3(Q^2)\big|_{\mathrm{QCD}}}, \label{s0Eqn} \\
\frac{\Phi_1(Q^2)\big|_{\mathrm{QCD}}}{\Phi_2(Q^2)\big|_{\mathrm{QCD}}} &= \frac{\Phi_1(Q^2)\big|_{\mathrm{HAD}}}{\Phi_2(Q^2)\big|_{\mathrm{HAD}}}, \label{widthEqn} \\
\Phi_1(Q^2)\big|_{\mathrm{QCD}} &= \Phi_1(Q^2)\big|_{\mathrm{HAD}}, \label{couplingEqn}
\end{align}
\end{subequations}
and  
\begin{equation}
\frac{\Phi_1(\tilde{Q}^2)\big|_{\mathrm{QCD}}}{\Phi_2(\tilde{Q}^2)\big|_{\mathrm{QCD}}} = M_V^2 + \tilde{Q}^2.
\label{QCDsumrulesmass}
\end{equation}
It can be observed that Eq.~(\ref{QCDsumrulesmass}) is the zero-width limit of Eq.~(\ref{widthEqn}), which motivates the introduction of two distinct momentum parameters, $Q^2$ and $\tilde{Q}^2$. The procedure for applying the Hilbert moment QCD sum rules to determine the effect of an external source on the hadronic state proceeds in two steps:

\begin{itemize}
    \item The first step consists of solving the Hilbert moment QCD sum rules in the vacuum. Using the experimentally measured vacuum hadronic width together with a chosen value of $Q^2$, we determine the vacuum values of the continuum threshold, the hadronic mass, and the coupling. Additionally, it is necessary to compute $\tilde{Q}^2$, so that Eq.~(\ref{QCDsumrulesmass}) is satisfied in the vacuum. 
    \item The second step involves solving the Hilbert moment QCD sum rules in the presence of an external source. To ensure the continuity of the hadronic parameters as the system transitions between the presence and absence of the external source, Eqs.~(\ref{QCDsumrules}) and Eq.~(\ref{QCDsumrulesmass}) will use the same $Q^2$ and $\tilde{Q}^2$ values, respectively, as those used in the vacuum. It is then a matter of calculating the hadronic parameters. Specifically, Eq.~(\ref{s0Eqn}) is used to obtain the continuum threshold parameter, Eq.~(\ref{QCDsumrulesmass}) is used to obtain the hadronic mass, Eq.~(\ref{widthEqn}) is used to obtain the width, and Eq.~(\ref{couplingEqn}) is used to obtain the coupling. These quantities must be computed in the order indicated.
\end{itemize}

The methodology of the Hilbert moment QCD sum rules has been successfully applied to heavy-quark systems in vacuum and at finite temperature~\cite{Dominguez:2018zzi,Dominguez:2009mk}. In the next section, we extend it to the case in which the system is embedded in a static, uniform magnetic field, leading to new corrections on the QCD side and potential shifts in the hadronic parameters that allow us to probe magnetic catalysis in the heavy-quark sector.

\section{\label{sec3} Charmonium in the presence of a magnetic field}

In order to incorporate the magnetic effects in this work, we rewrite the Hilbert moment QCD sum rules by including the new contributions that arise when the magnetic field is present. The starting point is to recognize that the definition of the Hilbert moments acquires a magnetic dependence. In the QCD sector, we obtain the magnetic Hilbert moment, which is Eq.~(\ref{QCDhilbertmoment}) with magnetic effects included. This is given by
\begin{equation}
    \varphi_N(Q^2,B)\big|_{\mathrm{QCD}} = \frac{(-1)^N}{N!} \left( \frac{d}{dQ^2} \right)^N \Pi(-Q^2,B)\big|_{\mathrm{QCD}},
    \label{QCDmagneticHilbertmoment}
\end{equation}
where the scalar function $\Pi$ is extracted from the two-point correlation function, which now depends on the magnetic field. Taking into account the vacuum perturbative and non-perturbative contributions, as well as the perturbative magnetic field contributions, the QCD magnetic Hilbert moment can be written as
\begin{align} \varphi_N(Q^2,B)\big|_{\mathrm{QCD}}&=\varphi_N(Q^2)\big|_{\mathrm{QCD}}+\Delta\varphi(Q^2,B)|_{\mathrm{QCD}} \nonumber \\
&=\varphi_N(Q^2)|_{\mathrm{pQCD}}+\varphi_N(Q^2)|_{\mathrm{npQCD}}\nonumber \\
&+\Delta\varphi_N(Q^2,B)|_{\mathrm{pQCD}},
\end{align}
where $\Delta\varphi(Q^2,B)$ represents the additional contribution induced by the finite external magnetic field.

In the hadronic sector, the continuum threshold parameter, as well as the $J/\psi$'s mass, width, and coupling, acquire dependence on the magnetic field strength, encapsulated by the hadronic magnetic Hilbert moment. This is a generalization of Eq.~(\ref{HilbertmomentsHAD}), written as
\begin{equation}
    \varphi_N(Q^2,B)\big|_{\mathrm{HAD}}= \frac{1}{\pi}\int_0^{\infty} \frac{\text{Im}\,\Pi(s,B)\big|_{\mathrm{HAD}}}{(s+Q^2)^{N+1}} ds,
\end{equation}
where the hadronic spectral function is such that below threshold, it is given by the magnetic field-dependent Breit-Wigner form, while above threshold, it is written using perturbative QCD, which also carries magnetic field strength dependency
\begin{align}
\text{Im}\,\Pi(s,B)\big|_{\mathrm{HAD}}&=\frac{2f_V^2(B)M_V(B)\Gamma_V(B)}{(s-M_V^2(B))^2+M_V^2(B)\Gamma_V^2(B)}\nonumber \\
&\times \Theta(s_0(B)-s) \nonumber \\
&+ \mathrm{Im} \, \Pi(s,B)\big|_{\mathrm{PQCD}} \Theta(s - s_0(B)).
    \label{BrietWignerMag}
\end{align} 
The transition from these Hilbert moments to their subtracted counterparts is akin to the procedure described for the vacuum. The subtraction term in Eq.~(\ref{subtractionTerm}) becomes
\begin{equation}
    I_N(Q^2,B)\equiv\frac{1}{\pi} \int_{s_0(B)}^{\infty} \frac{\mathrm{Im} \, \Pi(s,B)\big|_{\mathrm{HAD}}}{(s + Q^2)^{N+1}} ds,
\end{equation}
which gives rise to the subtracted magnetic Hilbert moments
\begin{equation}
\Phi_N(Q^2,B)|_{\mathrm{QCD}} \equiv \varphi_N (Q^2,B)|_{\mathrm{QCD}} - I_N(Q^2,B), 
\end{equation}
as well as
\begin{align}
 \Phi_N(Q^2,B)|_{\mathrm{HAD}}&\equiv\varphi_N (Q^2,B)|_{\mathrm{HAD}} - I_N(Q^2,B) \nonumber \\
 &= \frac{1}{\pi} \int_{0}^{s_0(B)} \frac{\mathrm{Im} \, \Pi(s,B)\big|_{\mathrm{HAD}}}{(s + Q^2)^{N+1}} ds.   
\end{align}
Quark-hadron duality allows us to write
\begin{equation}
   \Phi_N (Q^2,B)\big|_{\mathrm{QCD}}=\Phi_N(Q^2,B)\big|_{\mathrm{HAD}},
\end{equation}
which forms the basis for extracting the magnetic field dependence of the $J/\psi$'s parameters. 

At this stage, we have almost all the necessary ingredients to determine the magnetic behavior of the hadronic parameters $s_0$, $M_V$,  $\Gamma_V$, and $f_V$, through the solution of the Hilbert moment QCD sum rules in Eqs.~(\ref{QCDsumrules}) and~(\ref{QCDsumrulesmass}). The only missing component is the explicit expression for the Hilbert moments on the QCD side. This sector comprises two contributions: the perturbative and non-perturbative ones. 

For the perturbative QCD contribution, the two-point correlation function is given by
\begin{equation}
    \Pi_{\mu\nu}(q,B)\big|_{\mathrm{pQCD}}=\Pi_{\mu\nu}^0(q)\big|_{\mathrm{pQCD}}+\Delta \Pi_{\mu\nu}(q,B)\big|_{\mathrm{pQCD}},
    \label{perturbativecorrelationfunctionwithB}
\end{equation}
where the first term is the vacuum contribution, and the second term encodes the magnetic corrections.

To proceed, we include the magnetic field contribution. When a constant and homogeneous magnetic field $\vec{B}=B\hat{z}$ is present, Lorentz invariance is explicitly broken, making necessary to distinguish between directions parallel and perpendicular to the field. In this case, the momentum and the metric tensor are decomposed as
\begin{align}
q^\mu &= q^\mu_{\perp} + q^\mu_{\parallel}, \\
g^{\mu\nu} &= g^{\mu\nu}_{\perp} + g^{\mu\nu}_{\parallel},
\end{align}
with the components defined as
\begin{align}
q^\mu_{\perp} &= (0, q^1, q^2, 0), \nonumber \\
q^\mu_{\parallel} &= (q^0, 0, 0, q^3),
\end{align}
and
\begin{align}
g^{\mu\nu}_{\perp} &= \text{diag}(0, -1, -1, 0), \nonumber \\
g^{\mu\nu}_{\parallel} &= \text{diag}(1, 0, 0, -1).
\end{align}
This decomposition implies that $q^2 =q^2_{\perp}+ q^2_{\parallel}$. 

The charm quark propagator is modified in the presence of the magnetic field and receives corrections that can be organized as a power series in $|q_fB|$. Up to second order, the translationally invariant propagator in momentum space reads
\begin{align}
S_B(k) &= \frac{i}{k^2-m^2}-|q_fB| \frac{\gamma^1 \gamma^2 (\slashed{k}_{\parallel} + m)}{(k^2 - m^2)^2} \nonumber \\
&+2i |q_fB|^2 \frac{k^2_{\perp} (\slashed{k}_{\parallel} + m) - \slashed{k}_{\perp}(k^2_{\parallel} - m^2)}{(k^2 - m^2)^4},
\label{propagatorcharm}
\end{align}
where $q_f$ and $m$ are the electric charge and mass of the charm quark, respectively. This expression corresponds to the weak-field approximation, valid in the regime $|q_fB|\ll m^2$. Throughout this work, we use the vacuum charm quark mass $m^2=1.69 \ \text{GeV}^2$. 

\begin{figure}[t]
    \centering
    \includegraphics[scale=0.19]{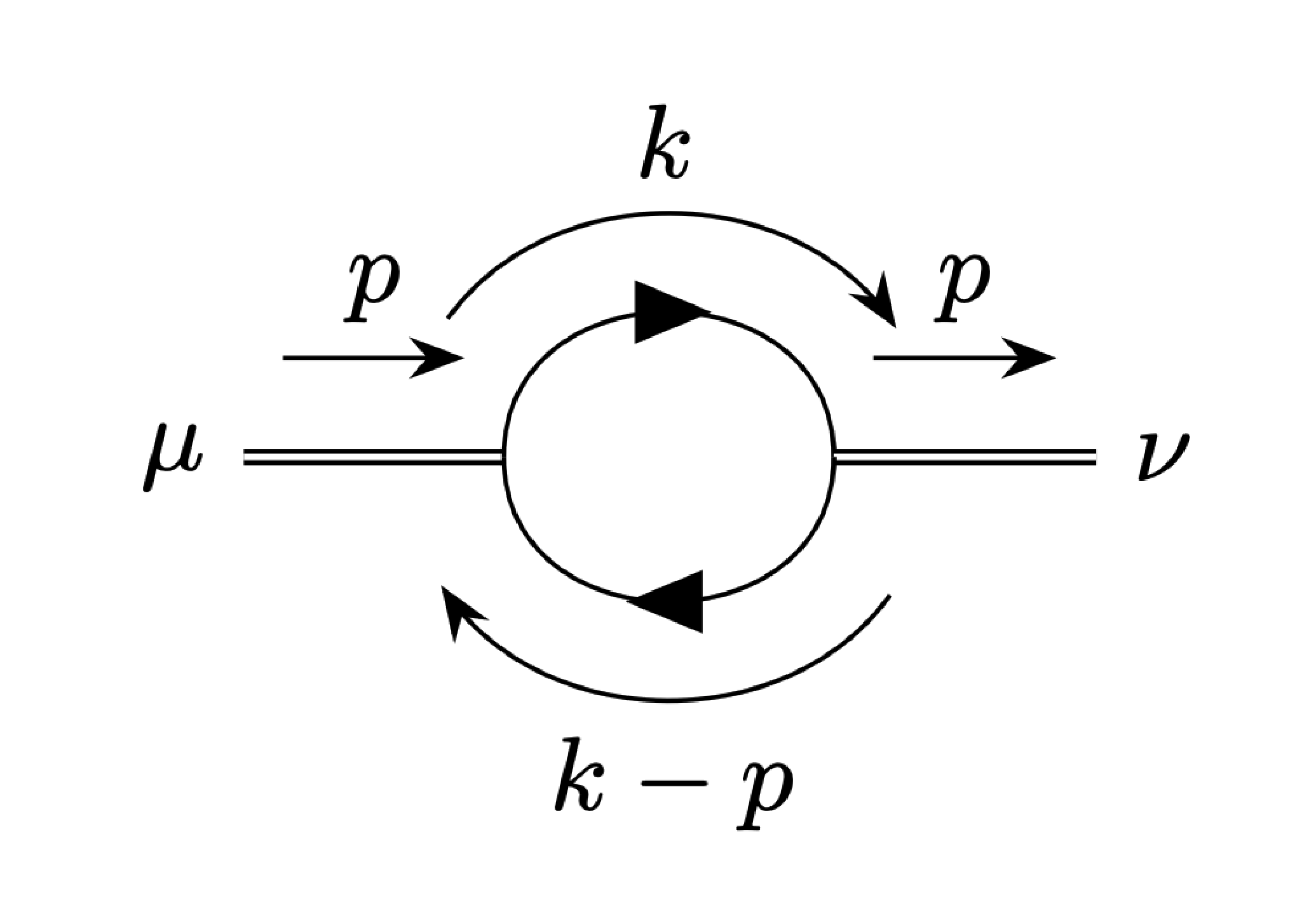}
    \caption{One-loop diagram representing the two-point correlation function in the charmonium vector channel.}
    \label{fig1}
\end{figure}

At one-loop order, the two-point correlation function in the vector charmonium channel is obtained from the Feynman diagram shown in Fig.~\ref{fig1}, where the loop is formed by a charm-anticharm pair. The expression for the full two-point vector correlator is
\begin{equation}
    \Pi^{\mu\nu}(q,B)\big|_\mathrm{pQCD}=-iN_c\int d^4x \, e^{i q \cdot x}\mathrm{Tr}\{\gamma^\mu S_B(x)\gamma^\nu S_B(-x)\}
\end{equation}
where $S_B(x)$ is the Fourier transform of the momentum space magnetic propagator in Eq.~(\ref{propagatorcharm}) and $N_c$ is the number of colors. Gauge invariance requires the correlation function to be transverse. However, the breaking of Lorentz symmetry due to the external magnetic field causes the tensor structure to split into three independent transverse components, which can be written as (see Ref.~\cite{Ayala:2020wzl,Hattori:2012je,Karmakar:2018aig})
\begin{align}
\Delta \Pi^{\mu\nu}(q,B)\big|_\mathrm{pQCD}&=X_1(q^2,q_\perp^2,B)Q^{\mu\nu}_\parallel+X_2(q^2,q_\perp^2,B)Q^{\mu\nu}_\perp\nonumber \\
&+X_3(q^2,q_\perp^2,B)Q^{\mu\nu}_0,
\label{Pipurelymagnetic}
\end{align}
where
\begin{align}
    Q^{\mu\nu}_\parallel&=g^{\mu\nu}_\parallel-\frac{q^\mu_\parallel q^\nu_\parallel}{q^2_\parallel}, \nonumber \\
    Q^{\mu\nu}_\perp&=g^{\mu\nu}_\perp-\frac{q^\mu_\perp q^\nu_\perp}{q^2_\perp}, \nonumber \\
    Q^{\mu\nu}_0&=g^{\mu\nu}-\frac{q^\mu q^\nu}{q^2}-Q^{\mu\nu}_\parallel-Q^{\mu\nu}_\perp.
\end{align}
The functions $X_i$ encode the magnetic field corrections up to second order in $|q_fB|$. Their explicit expressions, proportional to those computed in Ref.\cite{Ayala:2020wzl}, are shown in Appendix~\ref{AppendixA}.

The vacuum contribution to the two-point correlation function is
\begin{align}
    \Pi^0_{\mu\nu}(q)\big|_\mathrm{pQCD}&=-\left(g_{\mu\nu}-\frac{q_\mu q_\nu}{q^2}\right)\frac{3 q^2}{2\pi^2}\int_0^1dx \ x(1-x)\nonumber \\
    &\times \log\left(\frac{\mu^2}{m^2-x(1-x)q^2}\right),
    \label{Pivacuum}
\end{align}
where $\mu$ is the renormalization scale and the $\overline{MS}$ subtraction scheme is used. Thus, Eq.~(\ref{perturbativecorrelationfunctionwithB}) is fully specified by substituting Eqs.~(\ref{Pipurelymagnetic})-(\ref{Pivacuum}). 

On the other hand, the non-perturbative contribution to the Hilbert moments reads~\cite{Colangelo:2000dp,Dominguez:2009mk}
\begin{align}
    \varphi_N\big|_\mathrm{npQCD}(Q^2,B)&=-\frac{1}{3}\frac{2^N}{(4m^2)^{N+2}}\frac{N}{(1+\xi)^{N+2}}\nonumber \\
    &\times \mathrm{F}\bigg(N+2,-\frac{1}{2},N+\frac{7}{2},\delta\bigg) \nonumber \\
    &\times \frac{(N+1)^2(N+2)(N+3)(N-1)!}{(2N+5)(2N+3)!!}\nonumber \\
    &\times \bigg\langle\frac{\alpha_s G^2}{\pi}\bigg\rangle,
    \label{nonperturbativeHilbertmomentsB}
\end{align}
where $\xi= Q^2/4m^2$, $\delta= \xi/1+\xi$, $\mathrm{F}(a,b,c,z)$ is the hypergeometric function, and $\big\langle\frac{\alpha_s G^2}{\pi}\big\rangle$ is the gluon condensate.

We are now in a position to solve the Hilbert moment QCD sum rules, Eqs.~(\ref{QCDsumrules}) and~(\ref{QCDsumrulesmass}), as all the necessary ingredients have been identified and are now available for use. It is important to emphasize that, throughout this work, we adopt the kinematic choice $q^2_{\perp} = 0$, so that $q^2 = q^2_{\parallel} \equiv -Q^2$. This simplifies all expressions and defines a single scale for the Hilbert moments. Such a kinematic setup directly affects the perturbative QCD sector. 

\begin{figure}[t]
    \centering
    \includegraphics[scale=0.58]{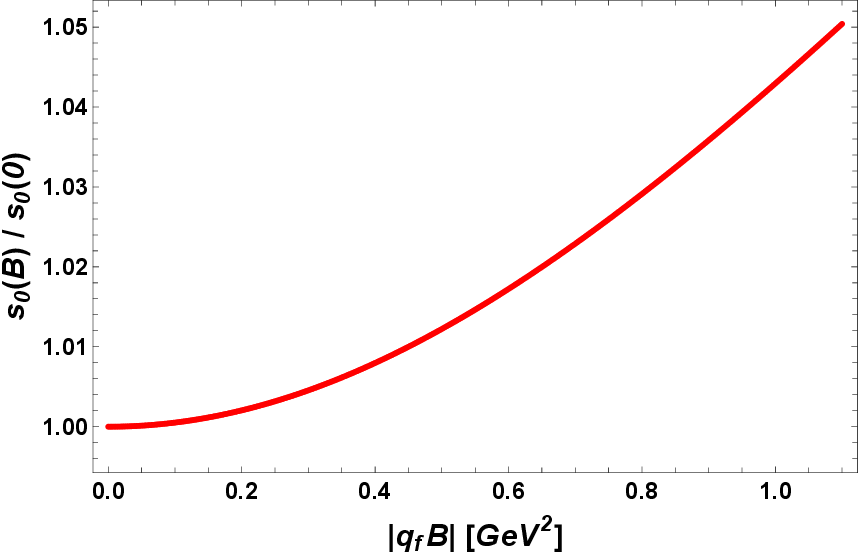}
    \caption{Squared energy pQCD threshold, $s_0$, normalized to its $|q_fB|=0$ value, as a function of the magnetic field strength.}
    \label{fig2}
\end{figure}

Moreover, since the magnetic field breaks Lorentz invariance and modifies the tensor structure, we choose to project the correlation function onto the full four-dimensional transverse tensor $g_{\mu\nu}-\frac{q_\mu q_\nu}{q^2}$. This implies that the magnetic corrections to the two-point correlation function are evaluated in this projected structure. Therefore, Eq.~(\ref{perturbativecorrelationfunctionwithB}) becomes
\begin{align}
    \Pi_{\mu\nu}(q,B)\big|_\mathrm{pQCD}&=-\left(g_{\mu\nu}-\frac{q_\mu q_\nu}{q^2}\right)\Bigg \{ \frac{3 q_\parallel^2}{2\pi^2}\int_0^1dx \ x(1-x)\nonumber \\ 
    &\times \log\left(\frac{\mu^2}{m^2-x(1-x)q_\parallel^2}\right)\nonumber \\
    &-\frac{|q_fB|^2}{3\pi^2}\Bigg[ \frac{3}{q_\parallel^2}+\frac{2(q_\parallel^2-6m^2)}{(q_\parallel^2)^{3/2}\sqrt{4m^2-q_\parallel^2}}\nonumber \\
    &\times \arctan\left(\sqrt{\frac{q_\parallel^2}{4m^2-q_\parallel^2}}\right)\Bigg]\Bigg\}.
    \label{PipQCDqparallel}
\end{align}

\section{\label{sec4} Results}

\begin{figure}[t]
    \centering
    \includegraphics[scale=0.58]{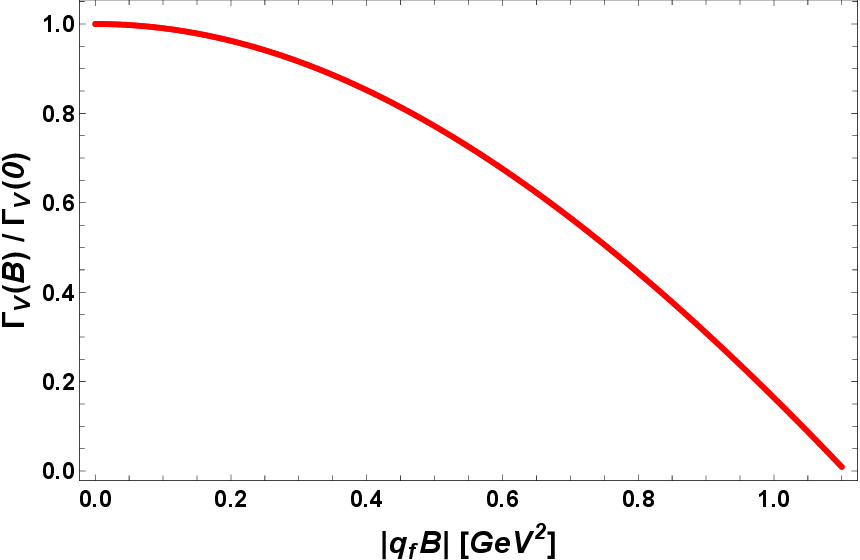}
    \caption{$J/\psi$ resonance width, $\Gamma_V$, normalized to its $|q_fB|=0$ value, as a function of the magnetic field strength.}
    \label{fig3}
\end{figure}

\begin{figure}[t]
    \centering
    \includegraphics[scale=0.58]{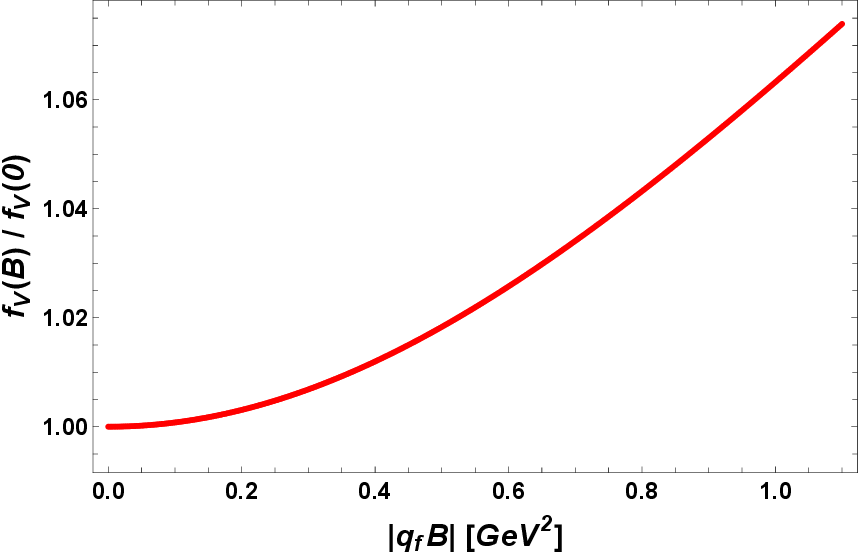}
    \caption{$J/\psi$ resonance decay constant, $f_V$, normalized to its $|q_fB|=0$ value, as a function of the magnetic field strength.}
    \label{fig4}
\end{figure}

\begin{figure}[h]
    \centering
    \includegraphics[scale=0.58]{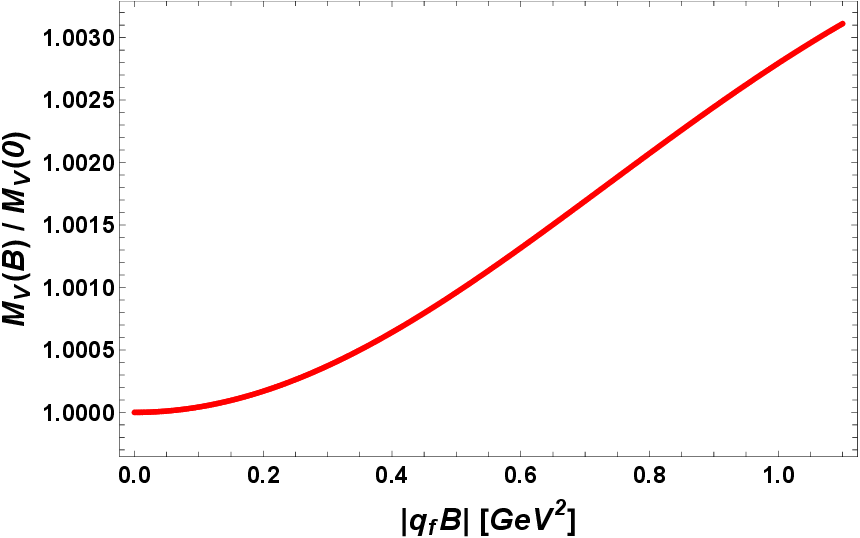}
    \caption{$J/\psi$ resonance mass, $M_V$, normalized to its $|q_fB|=0$ value, as a function of the magnetic field strength.}
    \label{fig5}
\end{figure}

We are now ready to solve and present the results of the Hilbert moment QCD sum rules. Recall that the magnetic behavior of the hadronic parameters $s_0$, $\Gamma_V$, and $f_V$ are the important unknowns to be extracted, as they are the indicators of the system tending towards or away from confinement. The first step is to determine the values of $s_0$, $M_V$, and $f_V$ in vacuum, as well as the value of $\tilde{Q}^2$, by solving the Hilbert moment QCD sum rules in the absence of a magnetic field. This requires specifying the charm quark mass $m$, the gluon condensate $\langle \frac{\alpha_s G^2}{\pi} \rangle$, the width of the resonance $\Gamma_V$, and a value for $Q^2$. In this work, we use $m = 1.3$ GeV, $\langle \frac{\alpha_s G^2}{\pi} \rangle = 0.06 \, \mathrm{GeV}^4$, $\Gamma_V = 93.2$ KeV, and $Q^2 = 10 \ \text{GeV}^2$. Solving the Hilbert moment QCD sum rules in vacuum yields the values: $s_0(B = 0) = 11.64 \ \text{GeV}^2$, $M_V(B = 0) = 3.09$ GeV, $f_V(B = 0) = 0.20$ GeV, and $\tilde{Q}^2 = 11.20 \ \text{GeV}^2$. Once all physical quantities in vacuum are fixed, we proceed to solve the Hilbert moment QCD sum rules in the presence of a magnetic field. The results are presented in Figs.~\ref{fig2}–\ref{fig5}.

Figure~\ref{fig2} shows the normalized phenomenological continuum threshold \( s_0 \) as a function of the magnetic field strength \( |q_f B| \). A monotonic increase is observed as the field becomes stronger, indicating that the onset of the perturbative QCD contribution shifts to higher energy scales. In Fig.~\ref{fig3}, we show the behavior of the normalized $J/\psi$ resonance width \( \Gamma_V \). We observe a significant and monotonic suppression as the magnetic field strength increases, eventually approaching a nearly vanishing value. This indicates that the resonance becomes increasingly narrow in the presence of a strong magnetic field. Figure~\ref{fig4} displays the normalized $J/\psi$ decay constant \( f_V \) as a function of the magnetic field. The decay constant exhibits a monotonic increase with increasing field strength, reflecting an enhanced coupling of the vector state to the electromagnetic current. Finally, in Fig.~\ref{fig5}, we show the normalized $J/\psi$ mass \( M_V \) as a function of \( |q_f B| \). A slight increasing trend is observed; however, it is practically negligible, indicating that the vector meson mass remains essentially stable under the influence of the magnetic field.

The combined behavior observed in the hadronic parameters as functions of the magnetic field strength leads to a consistent physical interpretation. The shift of the continuum threshold $s_0$ towards higher values, the suppression of the resonance width $\Gamma_V$, the monotonic increase of the decay constant $f_V$, and the almost constant, though slightly increasing, behavior of the $J/\psi$ mass all point towards an enlargement of the hadronic region as the magnetic field increases. The decrease in the resonance width implies an increase in the mean lifetime of the state, suggesting that the $J/\psi$ becomes more stable in the presence of a strong magnetic field. Simultaneously, the increase in the decay constant indicates a stronger coupling to the vector current, reflecting a more intense interaction strength.

These features are in line with the phenomenon known as \textit{Magnetic Catalysis}, typically understood as the strengthening of hadronic states in the presence of an external magnetic field. Although magnetic catalysis was originally formulated in the context of light quarks through the study of chiral symmetry restoration in QCD, such a framework is not applicable to heavy quark systems. Nevertheless, the magnetic field interacts purely through the electromagnetic sector and does not distinguish between flavor or color degrees of freedom, allowing its effects to be studied in the heavy quark sector as well.

By employing QCD sum rules and focusing on the phenomenological parameter $s_0$, which signals the onset of the perturbative QCD regime, we provide a novel and consistent analysis of magnetic field effects in heavy quark systems. Our results corroborate the manifestation of Magnetic Catalysis in the charmonium sector. The present study establishes a robust framework to explore magnetic field effects on hadronic parameters in the heavy quark sector using Hilbert moment QCD sum rules. While our focus has been on the vector channel of charmonium, the methodology is general and can be readily extended to other channels, such as pseudoscalar or axial-vector, and even to bottomonium states. These generalizations would allow for a broader understanding of how strong magnetic fields influence heavy quarkonia, opening the door to a systematic characterization of magnetic catalysis across different quark sectors.

\begin{acknowledgments}
Support for this work was received in part by the Secretaria de Ciencia, Humanidades, Tecnología e Innovación Grants No. CF-2023-G-433 and No. CBF-2025-G-1718. LAH acknowledges support from the DAI UAM PIPAIR 2024 project under Grant No. TR2024-800-00744. CAD and MK acknowledges support from the National Research Foundation (South Africa).
\end{acknowledgments}

\bibliography{mybibliography}

\appendix
\section{Magnetic scalar functions of the two-point correlation function \label{AppendixA}}

We explicitly write the coefficients corresponding to the three independent transverse tensor structures that arise in the presence of a finite magnetic field. These coefficients are given by

\begin{widetext}
\begin{align}
    X_1&=\frac{|q_fB|^2}{(q^2)^\frac{5}{2}(4m^2-q^2)^\frac{5}{2}\pi^2}\Bigg\{ \sqrt{4 m^2 - q^2} \sqrt{q^2} \Bigl[-6 m^4    q_{\perp}^2 + q^4 (q^2 - q_{\perp}^2) + m^2  q^2 (-4 m^2 + 7 q_{\perp}^2)\Bigr] \nonumber \\
   &+2 \Bigl[8 m^4   q^2 (q^2 - 2 q_{\perp}^2) + q^6 (q^2 - q_{\perp}^2) + 12 m^6  q_{\perp}^2 + 
   m^2 (-6 q^6 + 8 q^4  q_{\perp}^2)\Bigr]  \arctan\left(\sqrt{\frac{q^2}{4m^2-q^2}}\right) \Bigg \} ,
\label{coefX1}
\end{align}
\begin{align}
    X_2&=\frac{|q_fB|^2}{2(q^2)^{\frac{5}{2}}(4 m^2 - q^2)^\frac{5}{2} \pi^2}\Bigg\{ \sqrt{4 m^2 - q^2}  (6m^2 -q^2) \sqrt{q^2} \Bigl[ 8 m^2   q^2  - 2 q^4 - 14 m^2  q_{\perp}^2 + 5 q^2   q_{\perp}^2 \Bigr] \nonumber \\
   &-4 \Bigl[ 2 m^2   q^4 (2 q^2 - 7 q_{\perp}^2) + 12 m^6 (4 q^2 - 7 q_{\perp}^2) + q^6  q_{\perp}^2 + 
 m^4 (-28 q^4 + 58 q^2   q_{\perp}^2)  \Bigr] \arctan\left(\sqrt{\frac{q^2}{4m^2-q^2}}\right) \Bigg \},
\label{coefX2}
\end{align}
\begin{align}
    X_3&=\frac{|q_fB|^2}{(q^2)^{\frac{5}{2}}(4 m^2 - q^2)^\frac{5}{2}\pi^2}\Bigg\{ \sqrt{4 m^2 - q^2}   \sqrt{q^2}  \Bigl[6 m^4 (4 q^2 - 5 q_{\perp}^2) + q^4 (q^2 - q_{\perp}^2)  +  m^2  q^2 (-10 m^2 + 13 q_{\perp}^2)\Bigr] \nonumber \\
    &-4 m^2 \Bigl[-2 m^2  q^2 (7 q^2 - 9 q_{\perp}^2) + 6 m^4 (4 q^2 - 5 q_{\perp}^2) + 
   q^4 (2 q^2 - 3 q_{\perp}^2)\Bigr] \arctan\left(\sqrt{\frac{q^2}{4m^2-q^2}}\right) \Bigg \}.
\label{coefX3}
\end{align}
\end{widetext}

\end{document}